# Smart Cities: The Hopes and Hypes

Sudhir K. Routray
Department of Electrical and Computer Engineering
Addis Ababa Science and Technology University, Kilinto
Addis Ababa, Ethiopia
Email: sudhir.routray@aastu.edu.et

Susanta K. Sarangi
Department of Electronics and Telecommunication Engineering
Institute of Technical Education and Research, Bhubaneswar
Bhubaneswar, India
Email: susanta.sarangi@gmail.com

Abhishek Javali
Department of Telecommunication Engineering
CMR Institute of Technology, Bangalore
Bangalore, India
Email: abhishek.j@cmrit.ac.in

*Abstract*—Smart cities are being planned for several advanced applications and services for the inhabitants. Smart cities initiative promise many new services which are not possible in the traditional city frameworks. In the smart city framework, the basic aim is to provide all the essential services through sensor based systems which does not need much human intervention. This system is designed to operate on its own in a self-organizing manner. Therefore, the hopes are really big from the smart cities to enhance the quality of lives and the economy. However, some of the promises in the smart cities are very much over hyped. In this article, we analyse the realities of the smart cities and their practical significances based on the technological aspects of these projects. We also address the false promises that are around which are just the hypes. We clarify these hypes with appropriate logical explanations.

## I. INTRODUCTION

With the arrival of the Internet of things (IoT), there were several new technological initiatives to make the cities better for their inhabitants. This is how the smart cities initiatives gather pace across the world. Initially, under the smart cities framework the basic services in the cities were planned to be streamlined. It this initiative, many advantages were found for the overall growth and development of the cities. The economies of the cities get big boost through the smart city initiatives. It helps in the businesses, provides all basic facilities, and improves the quality of living. In the first two decades of this smart city framework several new facilities have been witnessed. Several cities in the developed countries have already provided some of the proposed services through their IoT based networks. The hopes for the smart cities are really overwhelming. However, in the recent past we have also witnessed a large number of hypes around this concept. Several service providers in the cities and even the city governments are advertising the smart city initiatives as the complete solution for all the problems of the cities. Even, these types of overestimated issues are also found in several scientific publications. Considering the technological realities we find that these promises of the advertisements are far from the truth.

Since the arrival of the IoTs, smart cities are very much in the limelight. Several research works have been carried out on these topics and the recent researches are very much active on multiple aspects of the IoT based smart cities. In [1], IoT for smart cities have been studied considering all the main aspects of the smart applications. The authors have analyzed the quality of services and the performance delays that may arise in the processes. They evaluated the solutions available for urban IoT deployment under the smart cities initiatives. Majority of the solutions discussed in [1] are already standardized and ready for deployment. The devices proposed for the smart city deployment are also been designed by different companies. In [2], the basic concepts of smart cities have been defined in the modern urban contexts. Practical aspects such as the dimension of the smart cities, their performance indicators, and implementation related aspects have been addressed in this paper. In [3], the economic and social aspects of smart cities have been studied. The authors show that smart cities are great economic divers. They can help in the businesses and the economy will certainly benefit from the smart cities initiatives. There are several impacts on the society as well. The safety and security aspects in the societies will get better in the smart cities. In [4], the smart cities are considered as intelligent cities in which several common aspects of the cities will be self-organizing and autonomous. In fact, smart cities will be the combined total of human, environmental and artificial intelligence. Therefore, according to the authors it will not be wrong to call the smart cities as the intelligent cities. In [5], the urban planning and design procedures of smart cities have been discussed. Newly built smart cities can have the advanced architecture from the beginning. However, the existing cities will go through several changes at various structural levels to accommodate the smart services and functions. In [6], the citizens' perspectives of smart cities have been addressed with respect to the changes those are expected. The smart cities will have a large number of information and communication technology (ICT) based services. These ICT services will have their own architecture and infrastructure. Several of these ICT services will be very much interactive. These interactive and autonomous activities of the smart cities will change the social fabric to a large extent. In [7], the information related benefits of smart cities have been addressed from the practical ICT applications points of views. Several aspects of the cities such as traffic on the roads, crowd during the festivals, and safety related information can be predicted from the smart city data. This will be helpful for both the businesses and individuals. Smart cities are going to have a large number of sensors for collecting and sensing the data. In large cities millions of sensors will be deployed. These huge deployments need a large amount of energy to make the smart cities functional. In [8], green IoTs have been considered for large scale deployment of sensors for smart cities. There are several energy efficient versions of IoTs. One of them is narrowband IoT (NBIoT). This version is very much energy and other resource efficient. NBIoT can be a good option for the smart cities as it consumes a very small amount of power. Recent trends in smart cities have been described in [9]. Majority of the trends show the realities of the smart cities and the amount of money required to make them function. However, several aspects are still very much opaque. It is also found that a lot of hypes are being spread about the smart cities. There are big gaps between the hypes and realities. A comprehensive survey has been carried out on the IoT and its applications in [10]. Smart cities are one of the main applications proposed for IoTs. Modern smart cities are not possible without IoT. The computing and communication related requirements of the smart cities can only be fulfilled by the IoTs. Smart cities are going to have an IoT traffic which is very much heterogeneous as they will come from many different types of sensors and actuators. There is a need of analyzing this IoT traffic. In [11], a traffic classification mechanism has been proposed which will segregate the traffic based on their types. In [12], the security aspects of IoT have been considered. As IoTs are going to be ubiquitous and will deal with many important aspects of individuals and businesses, its security is essential. Quantum cryptography is a robust method and in [12] it has been proposed a potential solution for IoTs. In smart cities location based services will have a big demand due to the large city population and availability of a large number of businesses. In [13], IoT based localization services have been addressed. It is shown that IoT based localizations are more accurate than the cellular-based and other localization methods in use these days. In [14], sensing applications in smart cities have been analyzed in detail. Sensing is the first and the most basic requirement of IoTs. In smart cities all the applications are based on IoTs and their accuracies are needed for better performances in the smart cities. Smart cities are based on a lot of intelligence and thus it is also often called as knowledge cities. In [15], several aspects of the knowledge related aspects of smart cities have been analyzed. It is found that the knowledge acquired through the smart city infrastructure is very much reliable. In this paper, several success stories of knowledge-based analysis of smart cities' data have been presented. In [16], healthcare provision using NBIoT has been addressed. It is shown that NBIoT is suitable for healthcare services as it deals with low energy and low bandwidths. It is also very much cost effective and easy to deploy. Therefore, NBIoT based healthcare for smart cities are very much attractive in the long term. There are several misinformation and hypes around the smart cities concepts. In [17], the realities and the hypes are clarified through the real methods of technology forecasting. It is true that smart cities will boost the economy but they are the solution to all the problems the cities face today. In [18], the smart cities have been studied and the hypes and realities have been clarified. In this work too smart cities have been scrutinized for the hypes. It is found that the smart city projects are promising several things which are beyond their reach. In [19], the detailed performances and hypes about the smart cities have been reported. The study collected information from several smart cities and analyzed their realities. It was found from the study that several aspects of smart cities have been overrated and far from realities.

In this article, we address the main benefits of the smart cities for the individuals and the businesses. We also show the hopes of the future smart city developments and their impact on the cities. We identified several hypes around the smart cities those are in circulation these days. We explain the reason about the hypes.

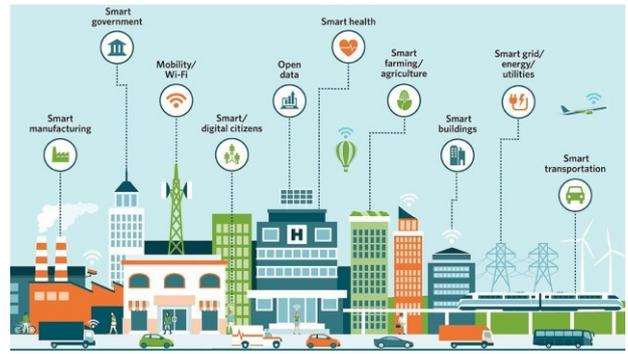

Fig. 1. The typical scenario of smart cities in which the basic resources and infrastructure are managed efficiently and effectively [1].

The reminder of this article is organized in four different sections. In Section II, we describe the fundamentals of the smart city initiatives. In Section III, we address the hopes and achievements of smart cities. In Section IV, we present the main hypes about the smart cities. In Section V, we conclude the paper with the main points.

## II. SMART CITY INITIATIVE

The idea of smart city was originated in the last couple of decades from the developments of information and communication tecnologies (ICT). Mainly due to the popularity of the mobile cellular technologies, the telecommunication infrastructure became ubiquitous around all the human settlement areas. It was also possible to provide several basic services for the cities using these ICT infrastructure. The arrival of the IoT made the things crystal clear. IoTs an be deployed over the cellular networks as value added services. These IoTs can be used for the smart city services. At the beginning, the idea was to provide the common facilities to the inhabitants of the cities. The basic facilities for the cities include, but are not limited to: continuous water supply; constant availability of electricity to each and every house in the city; lighting facilities in the cities from the sunset utill the sunrise; good transportation facilities; good policing and proper security for the citizens; appropriate cleaning and hygiene facilities in and around the cities; proper health facilities; emergency services; timely disaster management; and appropriate mechanisms to control and monitor the pollution levels in air, water and soil.

### A. Basic Principles of Smart City

Smart city is the concept of ICT enabled service provisioning in the cities to make the things efficient and autonomous. It was first proposed in the 1980s [18]. In the early years it did not get much success. Arrival of sensors and wireless sensor networks (WSNs) made the things better. In Europe and North America, several WSN based smart services such as crowd monitoring and smart policing were tried in different cities. However, the whole scenario got changed with the emergence of IoT. It was very soon realized in the 2010s that IoT based networks can provide much better services than the WSNs at a much lower cost. Thus several smart city initiatives were started in this decade to make the cities better place for living. IoTs are now the main ingredients of modern digital ecosystem. In Fig. 1, we show a typical scenario of smart cities.

TABLE I
DIFFERENT LEVELS OF QUALITIES AND RELIABILITIES USING IoTs IN SMART CITIES

| Services | Quality | Reliability |
|---|---|---|
| Street lighting | Good | Reliable |
| Accident emergency | Very good | Extremely reliable |
| Emergency healthcare | Very good | Extremely reliable |
| Fire service | Very good | Extremely reliable |
| Policing | Very good | Extremely reliable |
| Parking management | Good | Reliable |
| Transportation | Good | Extremely reliable |
| Water management | Very good | Reliable |
| Power management | Very good | Extremely reliable |
| Waste management | Good | Extremely reliable |
| Noise monitoring | Good | Reliable |
| Traffic monitoring | Good | Reliable |
| Air quality monitoring | Average | Reliable |
| Environmental issues | Average | Reliable |
| e-Governance | Average | Reliable |

## B. Why Smart City?

The main motivation for smart cities comes from the efficient use of resources. In the traditional approach, there is no robust control over the resource management. That is a big demerit for the large cities. Without a reliable and robust resource management the businesses and activities in the cities are affected very badly. In the modern times it is not acceptable to suffer losses due to the lack of proper facilities. In the smart city initiatives, all the basic and emergency services are allocated and managed through the IoTs. Through the sensors the ground realities are analyzed, and then through the actuators the services or information is provided. These services and assistances are provided much faster than the traditional services. Smart cities would make the tasks simpler for both the service providers and the citizens. In addition to all the above benefits, smart cities are going to provide a large number of economical benefits. The businesses, manufacturing, transportation, logistics, and retail management all are going to benefit from the efficient resource management of smart cities. Overall, we find the motivations for the smart cities are multi-faceted.

## III. HOPES AND ACHIEVEMENTS OF SMART CITIES

Since the beginning on the research on IoT based smart cities there are big hopes from these initiatives. As mentioned in Section II, the main aims of smart cities are to provide all the basic services to the inhabitants. In addition to the basic service provisioning, the quality of services have to be better than the traditional methods. There must be high guarantee in providing the emergency services such as accidental services, ambulance services, fire services and flood rehabilitation. These services become autonomous through the IoTs with some backup mechanism. The backups are essential to handle the situation in case of the gross failure of IoTs due to some natural disasters. In fact, the emergency services are made robust enough to sustain during the natural disasters. However, it is always a common practice to have a backup to make the services available all the time. The backup mechanisms are normally provided through more reliable channels such as wired networks with dedicated bandwidth for backup services.

The achievements of the smart cities have also been witnessed in several cities. In Western Europe, and several other developed countries, these smart city initiatives have been implemented through IoTs. They provide better services when compared with the traditional services through the old and inefficient methods. The services provided by the smart cities can be evaluated using their quality and reliability. In Table I, we provide some comparative figures for the performances in smart cities using the information available in some previous works [1] – [17]. We consider the quality to be 'average' if it is same as the traditional approach; 'good' if it is better, but become same as the traditional services occasionally; 'very good' if it is better than the traditional services almost all the instances. Similarly, for reliability we have two different levels depending on whether it is at the same level as the traditional systems or better than them. We compared the services at different level using the data given in [1]. The 'reliable' services are almost same or little better than the traditional services. However, the 'extremely reliable' one are almost always more reliable than the traditional services.

The services rendered by the smart cities are quite impressive when compared with the traditional approaches. It can be observed in Table I. This is one of the main motivations for going towards the smart city initiatives. It has also been observed that the smart cities help in the expansion of the economy to a large extent. The hopes from the smart cities are too many. In fact, the IoT based solutions are very much promising.

## IV. COMMON HYPES AROUND THE SMART CITIES

There are several hypes observed around the smart city concepts. These are the rumors and overestimations about the smart cities. Main reasons behind these hypes are due to the lack of knowledge of the smart city concepts, and sometimes intentional misguiding for some specific purpose. Either way, these hypes are the false information which must be avoided at the first place. The number of publications on smart cities is increasing almost exponentially [18]. With the rise in the number of publications, the hypes are also raising. In [19], it has been found the hypes around the smart cities are significant and the trends of these hypes are increasing very fast. Most of the hypes are related to the performance aspects of the smart cities [19]. The majority of the publications on the smart cities either overestimates or overrates the performances of the smart cities. In several cases, it has been witnessed that the overrating are intentional. In fact, it is very often due to the hyper optimism about the smart cities initiatives. It is also noteworthy that the academic publications contain the misinformation or the hyped information due to the lack of the availability of the real data. Some of the simulated results are very much out of reach of the real world. In fact, such simulators are designed by either the people who have limited knowledge about the IoTs for smart cities or the people who knowingly misguide by providing some wrong settings. These hypes will be reduced gradually with the better knowledge on the real performances of the IoTs in smart cities. However, the marketing related hypes are very much intentional. It is a breach of ethics and thus depends on the moral of the service providers. In fact, it will also come down gradually.

## A. The Common Instances of Hypes

In [19], several cases of common hypes around the smart cities have been presented. The most common cases are related to the very new concepts such as performances of 5G, autonomous systems, block chain based security systems, and several others. The smart cities are supposed to use all these technologies. But, the performance predictions about these new technologies are very much off the realities. In the cities where the smart city initiative is already in function also witness several hypes. In [19], it is explained that such hypes in the existing cases are very often due to the lack of budgeted amount with the city planners. That is why some of the performances are compromised at an inferior level than the initially planned ones. Another common instance of the failure to achieve the promised performance levels is the inability to fetch the revenue at the initial stages. For instance, a specific city needs fifty garbage trucks to collect the garbage every hour from all the locations. However, if the budget is not enough and they can only have forty garbage trucks then the performance will be certainly affected. Despite the accurate sensing, and timely information transfer, the garbage cannot be collected on time at some locations. Service providers of smart cities want to project their services aggressively. Their marketing gimmicks are very much there in the promotion of the smart cities through which they eye good revenues. Thus, it is noteworthy that the city governments must keep a vigil on the false promises of the service providers.

## B. The Ways to Convert the Hypes to Realities

Hypes are there due to several difficulties at different levels. Some of the hypes can be turned into realities with careful planning and proper deployment of the IoTs used for the smart cities. Smart cities are certainly huge projects which need a large sum of money for their successful deployment. All the cities cannot afford to have advanced smart systems. Therefore, the cities should always plan according to their abilities. However, they should always focus on the basic services, meaning the core services such as power management, water management, emergency services, and common facilities such as street light should be provisioned first. The other value added services such as free wireless Internet connection, and free transportation in certain localities are the secondary options. Private public partnerships are good solutions to overcome the funding crunches in the smart city projects. Right now, there are not too many IoT component manufacturers and service providers for smart cities. In the coming years, these trends are going to change by a large extent. The mature technology would provide better solutions and the hypes will be reduced significantly.

## V. CONCLUSIONS

In this paper, we presented the recent developments of smart cities which are very much dependent on the sensor based technologies. In fact, IoT is the main driving technology at the moment for the smart cities which can improve the quality of living in the cities. Smart cities are certainly good for the inhabitants and the environment. It makes the cities better place to live and flourish. However, the hypes too are on the rise. They must be avoided and the citizens should be provided the correct information. In fact, the marketing gimmicks of the service provider companies are very much there in the promotion of the smart cities. The belief that IoTs will make the cities smart is completely baseless. In fact, the onus is on the inhabitants. If they become smart they can use the technologies for their benefits and it would make the cities better. Overall, the smart inhabitants make a city smart, not the technology alone. Thus the real smart cities are the results of the joint collaboration between the conscious inhabitants and the efficient technologies. The hypes are the results of lack of knowledge on the smart cities, and the intentional marketing gimmicks by some service providers. These hypes around the smart cities initiatives will be reduced significantly over the next few years because people will understand the realities better.